\documentclass[11pt]{article}

\usepackage{epsfig,latexsym}
\def\aaa{angular momentum~}

\def\ba{\begin{eqnarray}}

\def\bbb{background~}
\def\bbbb{backgrounds~}
\def\be{\begin{equation}}
\def\bi{\bibitem}

\def\Bar {\overline}

\def\di{\partial}

\def\ea{\end{eqnarray}}
\def\ee{\end{equation}}

\def\em{energy-momentum~}
\def\ep{\epsilon}
\def\eq{\equiv~}
\def \EE{{\cal E}}

\def\fr{\frac}
 
 \def\gmn{g_{\mu\nu}}
\def\gr{general relativity~}

\def\ha{\frac{1}{2}~}

\def\ka{\kappa}

\def\kkk{Killing vector~}

\def\la {\lambda}
\def\lll{\left(}

\def\LLL{\left[}

\def\mn{\mu\nu}

\def\nl{\newline}

\def\nnn{\noindent}

\def\Om {\Omega}

\def\ra{\rightarrow}

\def\rrr{\right)}

\def\RRR {\right]}
\def\si{\sigma}
\def\sim{\simeq}
\def\sss{spacetime~}
\def\ssss{spacetimes~}
\def\sup{superpotential~}
\def\supp{superpotentials~}

\def\te{\theta}

\def\vf{\varphi}

\def\vs{\vskip 0.5 cm}

\def\1k{\fr{1}{\ka}}
\def\2k{\fr{1}{2\ka}}

\title{{\bf Comments on  conformal masses, \\asymptotic backgrounds
and conservation laws}}
\author{ Nathalie Deruelle \thanks {email:deruelle@ihes.fr}\\
{\it Institut des Hautes Etudes Scientifiques,}\\
{\it 35 Route de Chartres, 91440, Bures-sur-Yvette,
France}\\
\\ Joseph Katz\thanks{email:
jkatz@phys.huji.ac.il} 
\\{\it  The Racah Institute of Physics, Edmond Safra Campus}\\ {\it Givat Ram,
Jerusalem 91904, Israel}\\
\nl }

\begin{document}

\maketitle
\begin{abstract}
The ``conformal mass prescriptions" were used recently to calculate the mass of
\ssss in  higher dimensional and higher curvature theories of gravity. 
These  definitions are closely related to Komar  integrals for  spacetimes that are
conformally flat at great distances from the sources. We derive these relations 
without using the conformal infinity formalism.

\end{abstract}
 
 
\nnn {\it (i) Conformal mass and Komar integral}
\setlength{\baselineskip}{20pt plus2pt}

In a recent paper  Gibbons, Perry and Pope \cite {GPP} used  the prescription
given in Ashtekar and Magnon \cite{AM2} and  in Ashtekar and Das \cite{ADas} to
 calculate
the mass  of a rotating black hole in   ``anti-de Sitter  
 backgrounds   in higher dimensions [because] all the ambiguities that plague the
Komar prescription are avoided". Okuyama and Koga
\cite{OK} similarly generalized the  Ashtekar-Magnon-Das prescription  to
higher curvature  gravitational fields for different reasons, that is : ``the
conformal completion technique provides an intrinsic and background independent
definition of conserved quantities in an asymptotically AdS spacetime".   

Now,  Ashtekar and Magnon-Ashtekar \cite{AM} for  asymptotically flat
\ssss and Magnon \cite{Ma} for   asymptotically anti-de Sitter \ssss showed how to
derive  the ``conformal" mass and \aaa from the Komar integral
\cite{Ko}. The derivation was made  entirely in the conformal infinity formalism. Let
us do it here in the following way.

The Komar superpotential\footnote{Indices $\la,\mu, \nu, \rho,\cdots=0,1,2,3 $;
 the metric $\gmn(x^\lambda)$ has signature $+---$
and $g$ is its determinant. Multiplication by $\sqrt{-g}$ is indicated by an additional
hat,  $\hat X$, on a previously defined symbol like $X$.  $D$ is the covariant
derivative associated with the metric $g_{\mu\nu}$. The Levi-Civita symbol is
$\ep_{\mu\nu\rho\si}$ with $\ep_{0123}=1$. Square brackets
denote antisymmetrisation: $X_{[\mu\nu]}\eq{1\over2}(X_{\mu\nu}-X_{\nu\mu}$).
}$\hat J_K^{\mn}$  is an anti-symmetric tensor density defined as:
\be
\hat J_K^{\mn}\eq\1kD^{[\mu}\hat\xi^{\nu]}~~~{\rm with} ~~~\ka\eq\fr{8\pi
G}{c^4},
\label{1}
\ee 
 where $\xi^\mu$ is any vector field. Let $dS_{\mn}=\ha
\ep_{\mn\rho\si}dx^\rho\wedge  dx^\si$  be the  coordinate surface element of a
closed surface $S$. The Komar integral is 
\be
K\equiv\ha\oint_S \hat J_K^{\mn} dS_{\mn}.
\label{2}
\ee

The gist of the relation between Komar's integral and the
Ashtekar-Magnon  prescription, which involves the Weyl tensor, is  easily
understood as follows. Let $r^2 dS$ (with $dS=\sin\theta d\theta d\varphi$) be the
surface element of a sphere of radius $r$  in the asymptotically flat  background in
spherical coordinates.\footnote{We shall see below how to extend this argument to
an asymptotically anti-de Sitter spacetime.} Then 
$K$ is a finite quantity if its
integrant  is of the form
\be
K=\oint_{r\to\infty} f(r)\, r^2dS=\oint_{r\ra\infty }\lll   
\fr{a}{r^2}+\fr{b}{r^3}+\cdots
\rrr r^2 dS,
\label{3}
\ee
where $a, b,\cdots$ are functions  of $\te$ and $\vf$. Now $K$ may equally well be
written as
\be
K=\oint_{r\to\infty} \lll -\ha r\, \fr{\di f}{\di r}\rrr r^2 dS.
\label{4}
\ee
The covariant form of the integrant in (\ref{4}) {\it must} thus be proportional to
$-{1\over2}m^\rho D_\rho \hat J_K^{\mn}$, where $m^\rho$ is the unit vector in the
radial direction normal to the sphere. It can only be  linear in the curvature tensor
components that is, away from the sources where the Ricci tensor vanishes, in the
components of the Weyl tensor.   

Let us show that in an explicitly covariant way.
\vs


\nnn {\it (ii) Asymptotically flat \sss}
 \vs
  If  $\xi^\mu$   is a \kkk field, the covariant derivative of Komar's \sup  is related to
the Riemannian curvature tensor $R^{\mn}_{~~\rho\si}$ as follows :
\be
D_\rho  \hat J_K^{\mn}=\1k \,\hat g^{\si[\mu}D_{\rho\si}\xi^{\nu]}=-\1k\, \hat
R^{\mn}_{~~\rho\si}\xi^\si.
\label{5}
\ee       

 Let  $n^\mu$ and $m^\nu$  be two   orthogonal
normalized vectors orthogonal to $S$:
\be
g^{\mu\nu}n_\mu
n_\nu=1~~~,~~~g^{\mu\nu}m_\mu m_\nu=-1~~~{\rm and}~~~g^{\mu\nu}m_\mu
n_\nu=0.\label{6}
\ee
The coordinate surface element  $dS_{\mu\nu}$ may be written
as $dS_{\mu\nu}=-2n_{[\mu}m_{\nu]} d^2x$ where  $ d^2x$ is thus defined as 
$d^2x=n^\mu m^\nu dS_{\mu\nu}$. We therefore have, for any Killing vector field
$\xi^\mu$ :
\be
{1\over2}m^\rho (D_\rho \hat J_K^{\mu\nu}) dS_{\mu\nu}={1\over\kappa}\hat
R_{\mu\nu
\rho\lambda}\,\xi^\lambda\,n^\mu m^\nu m^\rho d^2x\, .
\label{7}
\ee
$\xi^\mu$ may only  be   an asymptotic \kkk of
time translations in which case (5) and (7) are only asymptotically true. In
asymptotically static and spherical coordinates $(t,r,\te,\vf)$ in which
$\xi^\lambda\ra (1,0,0,0)$, $n^\nu\ra(1,0,0,0)$, $m^\nu\ra(0,1,0,0)$, $\sqrt{-g}\to
r^2\sin\theta$, $d^2x=d\theta\, d\varphi$, $dS=\sin\te\, d\te\, d\vf$   and
$g_{00}\ra 1-2GM/(c^2r)$ (see e.g. \cite{MTW}),  a short calculation yields : 
\be
\lim_{r\ra\infty} \LLL -\ha r\oint_{r\ra\infty} \ha  m^\rho (D_\rho \hat
J_K^{\mu\nu}) dS_{\mu\nu}\RRR=\ha\oint \hat J_K^{\mn} dS_{\mn}= \ha Mc^2,
\label{8}
\ee
which, following (\ref{3}) and  (\ref{4}),  is the expected result connecting the Komar
integral and its derivative. Eq. (\ref{8})   shows as is well known that Komar's integral
yields half the expected value of the mass.\footnote{One may claim that this is a
matter of normalization. See, however, section {\it (iv)}.}

The connection between the integral of the radial derivative of   Komar's \sup
and the Ashtekar-Hansen \cite{AH} conformal mass is also very simply obtained. Since 
$\xi^\mu\ra n^\mu$, we have, from eq. (\ref{8}) with (\ref{7}):
\be
Mc^2=\lim_{r\ra\infty} \LLL - r\oint_{r\ra\infty}{1\over\kappa}  (m^\rho m^\si
R_{\mu\rho\si\nu})  n^\mu n^\nu r^2 dS\RRR.
\label{9}
\ee
Outside the sources, the Ricci tensor vanishes by virtue of   Einstein's
equations and 
$R^\la_{~\nu\rho\si}=C^\la_{~\nu\rho\si}$, where $C^\la_{~\nu\rho\si}$ is the Weyl
tensor. At conformal flat infinity the Ashtekar-Hansen conformal factor  
$\Om\to1/r^2$. We may therefore rewrite (\ref{9}) as follows: 
\be
Mc^2= -{1\over2} ~~ {c^4\over G} \oint_{r\ra\infty}  E_{\mu\nu} 
(\Omega^{-1}n^\mu n^\nu) {dS\over{4\pi}}  ~~~{\rm
in~which}~~~E_{\mu\nu}=\Om^\ha m^\rho  (\Omega^{-1}m^\si) C_{\mu\rho\si\nu}.
\label{10}
\ee
This expression is, with slightly different notations (in particular our  metric is
denoted without a hat), that given in  Ashtekar and Magnon \cite{AM} on page 796
where they relate the conformal mass to the Komar  integral in exactly the same way
but in a conformal infinity formulation.

\bigskip


\nnn {\it (iii) Asymptotically anti-de Sitter \sss}
 \vs

When spacetime is asymptotically anti-de Sitter, the Komar integral (\ref{2})
diverges. A way to remedy   this is  ``renormalization": one subtracts from $\hat
J_K^{\mn}$ the Komar \sup of the anti-de Sitter  \bbb $\Bar{\hat J_K^{\mn}}$ (see
\cite{Ma} or \cite{KBL}):
\be
\hat J^{\mn}_{K}-\Bar{\hat J^{\mn}_{K}}\eq\1k\ 
D^{[\mu}\hat\xi^{\nu]}-\1k\ \Bar{D^{[\mu}\hat\xi^{\nu]}}\,,
\label{11}
\ee 
where   barred quantities are defined in the anti-de Sitter background spacetime with
metric $\Bar {g}_{\mu\nu}\,(x^\lambda)$.\footnote{For a detailed definition of what is
meant by a background and the mapping on physical spacetime, see e.g. \cite{BK}. 
Let us emphasize that the background is defined in a coordinate independent way. For
instance,  for a Kerr \sss on an anti-de Sitter \bbb the \bbb metric is $\Bar
g_{\mn}=g_{\mn}(m=a=0)$ in whatever coordinates $g_{\mn}$  has been written. 
}

 If  $\xi^\mu$ and  $\Bar{\xi}^\mu$ are \kkk fields we have, as above:
\be
{1\over2}\left(m^\rho D_\rho \hat J_K^{\mu\nu} -\Bar{m^\rho
D_\rho \hat J_K^{\mu\nu}} \right)dS_{\mu\nu}={1\over\kappa}\left(\hat R_{\mu\nu
\rho\lambda}\,\xi^\lambda\,n^\mu m^\nu m^\rho-\Bar{\hat
R_{\mu\nu
\rho\lambda}\,\xi^\lambda\,n^\mu m^\nu m^\rho}\right) d^2x.
 \label{12}
\ee
The metrics $g_{\mu\nu}$ and $\Bar{g}_{\mu\nu}$  coincide at infinity and $\Bar{R}_{\mu\nu\rho\sigma}=-{1\over l^2}(\Bar{g}_{\rho\nu}\,
\Bar{g}_{\mu\sigma}-\Bar{g}_{\sigma\nu}\,\Bar{g}_{\mu\rho}$) where $l^2$ is the
characteristic scale of the anti-de Sitter spacetime. In static and spherically
symmetric coordinates  $(t,r,\te,\vf)$ :
\be 
d\Bar{s}^2=\Bar{g}_{\mu\nu}\,dx^\mu
dx^\nu=(1+r^2/l^2)dt^2-{dr^2\over1+r^2/l^2}-r^2(d\theta^2+
\sin^2\theta d\varphi^2)\,.
\label{13}
\ee

If $\xi^\mu\to\Bar{\xi}^\mu$ is the asymptotic \kkk of time translations we have :
\be
\xi^\la \to\Bar\xi^\la =(1,0,0,0)~~~,~~~n^\nu\to\Bar n^\nu\sim
\fr{l}{r}(1,0,0,0)~~~{\rm and}~~~ m^\nu\ra\Bar m^\nu\sim \fr{r}{l}(0,1,0,0).
\label{14}
\ee
As for the asymptotic behavior of the needed metric components 
$g_{\mu\nu}$, we have \cite{Ma}: 
\be
 g_{00}\ra \Bar g_{00}-\fr{2GM}{c^2r}~~~,~~~ g_{11}\to\Bar
g_{11}-\fr{2GM l^4}{c^2 r^5},
\label{15}
\ee
$S$ being again the sphere of radius $r\ra\infty$. One thus gets, see \cite{DKO}: 
\be
\ha\oint_{r\ra\infty} \left(\hat J_K^{\mn} -\Bar{\hat J_K^{\mn}}\right)dS_{\mn}=\ha
Mc^2.
\label{16}
\ee
Notice that the non-zero component of $m^\rho$ is not $1$ but $r/l$. Thus $m^\rho
D_\rho$  already contains the necessary factor $r$ which appears in (\ref{4}) but has
also an unnecessary factor
$l$ which we  remove by multiplication and we obtain therefore : 
\be
 -\ha l\oint_{r\ra\infty}{1\over2}\left(   m^\rho D_\rho \hat J_K^{\mu\nu}
-\Bar{m^\rho D_\rho \hat J_K^{\mu\nu}}\right) dS_{\mu\nu}= \ha Mc^2.
\label{17}
\ee
This is the expected result connecting the renormalized Komar integral (\ref{16})
and its derivatives (\ref{17}).

The connection between (\ref{17}) and the
Ashtekar-Magnon-Das conformal mass formula is again very simply obtained. We
have
\be
Mc^2= - l\oint_{r\ra\infty}{1\over\kappa}  
m^\rho m^\si (R_{\mu\rho\si\nu}-\Bar{R}_{\mu\rho\si\nu})  n^\mu \xi^\nu r^2 dS.
\label{18}
\ee
Outside the sources the Einstein equations are  $R_{\mu\nu}=3g_{\mu\nu}/l^2$ and
imply
$R_{\mu\rho\si\nu}=C_{\mu\rho\si\nu}-\fr{1}{l^2}(\gmn g_{
\rho\si}-g_{\mu\si} g_{\nu\rho} )$. Hence, at infinity,  
$(R_{\mu\rho\si\nu}-\Bar{R}_{\mu\rho\si\nu})\to C_{\mu\rho\si\nu}$ because there,
$\gmn=\Bar\gmn$ and
$\Bar C_{\mu\rho\si\nu}=0$. Thus (\ref{18})  can also be written as:
\be
Mc^2= -\fr{l}{8\pi} ~ \fr{c^4}{G} \oint_{r\ra\infty}   (m^\rho m^\si
C_{\mu\rho\si\nu})~ \xi^\mu n^\nu r^2\,dS \,.
\label{19}
\ee
This formula is the Ashtekar-Magnon-Das \cite{AM2} \cite{Ma} \cite{ADas} formula 
which reads,  in the (slightly adapted) notations of   \cite{OK}  :
\be
 Mc^2=-\fr{l}{ 8\pi}~\fr{c^4}{ G} \oint~~\hat\EE_{\mu\nu}~\xi^\mu\hat N^\nu
\sqrt{\hat\si }\, d^2x\quad\hbox{with}\quad
\hat\EE_{\mu\nu}=l^2\lim_{{\Om\ra\infty}}(\Om\,\hat M^\rho\hat
M^\sigma\,C_{\mu\rho\sigma\nu} ).
\label{20}
\ee
Here
$\sqrt{\hat\si}\,d^2x\equiv dS$,  $\hat N^\nu= l\,\xi^\nu\equiv r\, n^\nu$, with
$n^\nu$ defined in (14)~; the conformal factor $\Omega\ra 1/r$, $\hat
M^\rho\equiv (r/l)m^\rho$, with
$m^\rho$ defined in (14), so that $\hat\EE_{\mu\nu}\equiv r (m^\rho m^\si
C_{\mu\rho\si\nu})$.


\vs
\nnn {\it (iv) Comments}
 \vs

The Ashtekar-Hansen and Ashtekar-Magnon-Das global conservation laws are based
on the Penrose  formalism \cite{Pe} in which asymptotic regions are described by a
conformal \sss  with an appropriate conformal factor $\Om$  which depends on the
underlying asymptotic physical metric.

We saw that those prescriptions amount to computing the radial derivative
of Komar's integral $K$.  For
isolated stationary sources in  asymptotically flat spacetimes if $\xi^\mu$ is
the Killing vector of time translations $K=\ha Mc^2$ but if $\xi^\mu$ is a   Killing
vector of spatial rotations then $K =c J$ where
$J$ is the corresponding component of the angular momentum. This factor 2
discrepancy is the Komar anomalous factor \cite{Pe}.   (Ashtekar and Hansen formula is
normalized to yield the right mass.)
 In an asymptotically anti-de
Sitter 
\sss $K$ diverges. This is the Komar integral ``ambiguity" alluded to in
\cite{GPP}.  As we recalled, this divergence is  easily removed if instead of $K$
 we use \cite{KBL} $K-\Bar K$  
where $\Bar K$ is the Komar integral evaluated in the anti-de Sitter
``background". The anomalous factor 2 however remains. Finally, at null infinity
($u\equiv t-r$=const.) $K$ does not \cite{WT}  give  the Penrose \cite{Pe2}
mass  nor is
$dK/du$ the Bondi \cite{Bo} mass loss. Thus, the Komar integral does not
describe properly conserved quantities associated with asymptotic
symmetries\footnote{Conservation laws are associated with global \sss symmetries and
a variational principle. The Komar \sup is a construct with an arbitrary field 
 and is not related to   a variational principle (see however below) or to
global symmetries. The same is true for the Ashtekar et al. constructs.}. Notice also that
the Ashtekar-Magnon-Das conformal prescription  depends on second derivatives of the
metric and   requires thus more information on the background  than the Komar
integral. 

In spite of its limitations the Komar superpotential played an important role in the
development of conservation laws in general relativity and in the discovery of the
first law of black hole thermodynamics \cite{BCH}. It is also fair to say that the
conformal mass and conservation laws of Ashtekar et al. appeared   before the
limitations of Komar integrals became apparent and were corrected. 

The Komar \sup
\cite{Ka} can be derived from the  Hilbert Lagrangian  (the scalar curvature) 
using traditional Noether  identities.
It is a characteristic property of global conservation laws in general relativity
derived from   any Lagrangian by means of these  identities \cite{JS1}  that they
depend on a vector field $\xi^\mu$ and that conserved vector densities, say
 $\hat J^\mu$, are   divergences of anti-symmetric tensor densities or
superpotentials $~$:
  $\hat  J^\mu=\di_\nu \hat J^{\mu\nu}$, like the Komar superpotential $\hat
J_K^{\mu\nu}$. The asymptotic behavior of the gravitational field is also an essential
ingredient in the construction of globally conserved quantities. In most cases 
spacetimes become flat at great distances from the sources but   there has also been
great interest in asymptotically anti-de Sitter spacetimes.  In any
cases it is   useful to regard
\ssss at great distances from the sources as  backgrounds with a metric of their
own.  In global considerations we are not interested in how  \ssss are
mapped on \bbbb at finite distances. Backgrounds are just another  convenient
and {\it covariant} way to deal  with asymptotic limits. 

There are, however, well known difficulties with
\supp in that they are not uniquely defined. Recent work by Silva \cite{Si}, Julia and
Silva \cite{JS2} and Chen and Nester
\cite{CN1} \cite{CN2} based on ideas of Regge and Teitelboim
\cite{RT} lifted such ambiguities by imposing reasonable conditions that make $\hat
J^{\mu\nu}$ essentially unique. That  unique KBL superpotential\footnote{It has also
been shown
\cite{JS3} that the  criterion used to define the KBL \sup is indeed equivalent to the
covariant simplectic phase space methods of Wald \cite{Wa2}, Iyer and Wald   
\cite{IW},  Witten \cite{Wi} and Ashtekar, Bombelli and Reula \cite{ABR}.} \cite{Ka},
\cite{KBL}  is: 
\be
\hat J^{\mn}\eq  \hat J_K^{\mn}-\Bar{\hat 
J_K^{\mn}}+\1k\xi^{[\mu}\hat k^{\nu]}~~~{\rm with}~~~\hat  k^\nu\eq
\fr{1}{\sqrt{-g}}\Bar D_\mu(-gg^{\mn}).
\label{21}
\ee
According to Julia and Silva  \cite{JS2} this \sup is the only one possessing  all the
following properties. It is generally covariant  and can be computed in any coordinate
system. In cartesian coordinates of an asymptotically flat \sss it gives the ADM mass
formula
\cite{ADM} and in asymptotically anti-de Sitter \ssss it gives the AD mass \cite{AD}. It
gives the mass and \aaa as well as the Brown and Henneaux conformal charges
\cite{BH} with the right normalization in any dimensions $D\ge 3$. It can
also be used for any \kkk field of the background. It reproduces the Penrose mass
\cite{Pe} the Bondi mass loss \cite{BVM},  the Penrose linear momentum \cite {Pe}, the
Sachs linear momentum flux \cite{Sa} and the Penrose \cite{Pe2} and Dray and
Streubel \cite{DS} angular momentum at null infinity. One may add that it gives the
mass of a Kerr black hole in an anti-de Sitter background in $D$ dimensions \cite{DK}.
Finally the extension of the KBL prescription to Einstein-Gauss-Bonnet theory of
gravity
\cite{DKO} gives also the mass and angular momenta of rotating black holes on AdS
backgrounds in that theory \cite{DM}.

Of course, if one is interested in calculating conservation laws at spatial infinity
only,  either flat or AdS, one can as well use the ADM  superpotential
\cite{ADM} or  the Abbot and Deser one \cite{AD} or  the   Petrov and Katz one
\cite{PK}  or, in Einstein-Gauss-Bonnet theory, the Deser and Tekin one \cite{DTa},
\cite{DTb} or other covariant Hamiltonian formulations many of which are reviewed in
the recent
 paper of Szabados \cite{Sz}.

\vskip 1cm
\nnn{\bf Acknowledgements}
 
\nnn N.D. thanks the Hebrew University for hospitality and
financial support.

\vskip 1cm
\nnn{\bf Note added in proof~:} In this paper we limited ourselves to a comparison between the
Komar and Ashtekar et al. {\it masses}. It is important to note that equation (9) (or (10)) where
$n^\mu$ is replaced by $\xi^\nu_\phi=(0,0,0,1)$ gives also the right value of the Kerr {\it angular
momentum}~: $J=M a c$. Similarly equation (19) (or (20)) gives the right Kerr AdS mass {\it and}
angular momentum, as shown explicitely in \cite{Koga}. Hence defining the mass and angular
momentum by means of the {\it derivative} of the Komar superpotential (or, equivalently, as we
have shown, by means of  equations (10) and
(20)) cures the problem of the anomalous factor 2 which plagues the Komar integrals.
(Note however that Ashtekar Hansen and Das 
\cite{ADas} \cite{AH} use a different construct for the angular momentum.)
We are grateful to Jun-ichirou Koga for bringing our attention to that point.
\nnn

 \end{document}